\documentclass[preprint2,twoside]{hwo}

\usepackage{lipsum}

\input{hwo.h}

\begin{document}

\title{\textbf{\LARGE Deciphering The Launching of Multi-phase AGN-driven Outflows and Their (Spatially Resolved) Multi-scale Impact}}
\author {\textbf{\large Lulu Zhang,$^{1}$ Gagandeep Kaur,$^{2,3}$ Tianmu Gao,$^{4,5}$ \'{A}lvaro Labiano,$^6$ Erin K. S. Hicks,$^{7}$ Vivian~U,$^{8,9}$ Chris Packham,$^{1,10}$ Missagh Mehdipour,$^{11}$ Travis Fischer,$^{11}$ Thaisa Storchi Bergmann,$^{12}$ Namrata Roy,$^{13}$ Isabel M\'{a}rquez,$^{14}$ and Christiaan Boersma$^{15}$ }}
\affil{$^1$\small\it The University of Texas at San Antonio, San Antonio, Texas, USA; l.l.zhangastro@gmail.com}
\affil{$^2$\small\it Graz University of Technology, Rechbauerstraße 12, 8010 GRAZ, Austria}
\affil{$^3$\small\it Space Generation Advisory Council (SGAC), Schwarzenbergplatz 16, TOP 1 1010 Vienna, Austria}
\affil{$^4$\small\it Australian National University, Weston Creek, Australian Capital Territory, Australia}
\affil{$^5$\small\it ARC Centre of Excellence for All Sky Astrophysics in 3 Dimensions (ASTRO 3D)}
\affil{$^6$\small\it Telespazio UK for the European Space Agency, Camino Bajo del Castillo s/n, Villanueva de la Can\~{a}da, Spain}
\affil{$^7$\small\it University of Alaska Anchorage, Anchorage, Alaska, USA}
\affil{$^8$\small\it California Institute of Technology, Pasadena, California, USA}
\affil{$^9$\small\it University of California, Irvine, Irvine, California, USA}
\affil{$^{10}$\small\it National Astronomical Observatory of Japan, Mitaka, Tokyo, Japan}
\affil{$^{11}$\small\it Space Telescope Science Institute, Baltimore, Maryland, USA}
\affil{$^{12}$\small\it Instituto de Fisica da UFRGS, Porto Alegre, RS, Brasil}
\affil{$^{13}$\small\it Johns Hopkins University, Baltimore, Maryland, USA}
\affil{$^{14}$\small\it Instituto de Astrof\'{i}sica de Andaluc\'{i}a (IAA-CSIC), glorieta de la astronomía S/B, 18008 Granada, Spain}
\affil{$^{15}$\small\it NASA Ames Research Center, Moffett Field, California, USA}

\author{\footnotesize{\bf Endorsed by:}
Athina Meli (North Carolina A\&T State University), Sanskriti Das (Stanford University), Sylvain Veilleux (University of Maryland), Filippo D'Ammando (INAF-IRA Bologna), Eunjeong Lee (EisKosmos (CROASAEN), Inc.), Greg Bryan (Columbia University), Stefano Bianchi (Università degli Studi Roma Tre), Kate Rowlands (STScI), Sophia Flury (University of Edinburgh), Stephanie LaMassa (STScI), Evan Scannapieco (Arizona State University), Anton Koekemoer (STScI), Rebecca Davies (Swinburne University of Technology), Andreea Petric (STScI), and Patil Pallavi (Johns Hopkins University)}

\begin{abstract}
Beyond deepening our understanding of the formation, growth, and evolution of supermassive black holes, it is crucial to uncover the role of feeding and feedback processes from growing black holes (i.e., active galactic nucleus; AGN) in shaping the cosmic ecosystem. Such studies include understanding the dynamics of gas flows in the interstellar (ISM), circumgalactic (CGM), intracluster (ICM), and intergalactic media (IGM). As the output of a sub-group in Habitable Worlds Observatory (HWO) AGN Working Group, this Science Case Development Document (SCDD) proposes to use future HWO observations to solve the following questions. Which mechanism is dominant in triggering inflows/outflows through feedback? How is AGN activity triggered, and is it associated with circumnuclear star formation and what is the overall effect of AGN feedback on star formation (SF)? In AGN feedback, which mode is more influential and does AGN feedback operate similarly or differently in the local universe and at high redshift? To answer these questions, this SCDD proposes to use potential HWO observations as follows. Resolve and characterize the spatial distribution of ionized and cold/warm molecular gas, especially those in inflows/outflows; Explore the spatial coupling and potential stratification of multi-phase inflows/outflows on different physical scales and their resolved and global correlations with AGN and/or SF activities; Investigate whether corresponding outflows/jets induce shocks and/or fluctuations that trigger or suppress the formation of molecular clouds and hence new stars. Specifically, HWO's capabilities will enable us to achieve the above scientific goals while existing facilities lack the required combination of high-throughput ultraviolet (UV) and near-infrared (NIR) integral field unit (IFU) capabilities with simultaneously sufficient spatial resolution and sensitivity.
  \\
  \\
\end{abstract}

\vspace{2cm}

\section{Science Goal: What Role Do Feeding and Feedback of Growing Black Holes Play in Galaxy Evolution?}\label{sec1}

\begin{figure}[!ht]
\centering
\includegraphics[width=1\columnwidth]{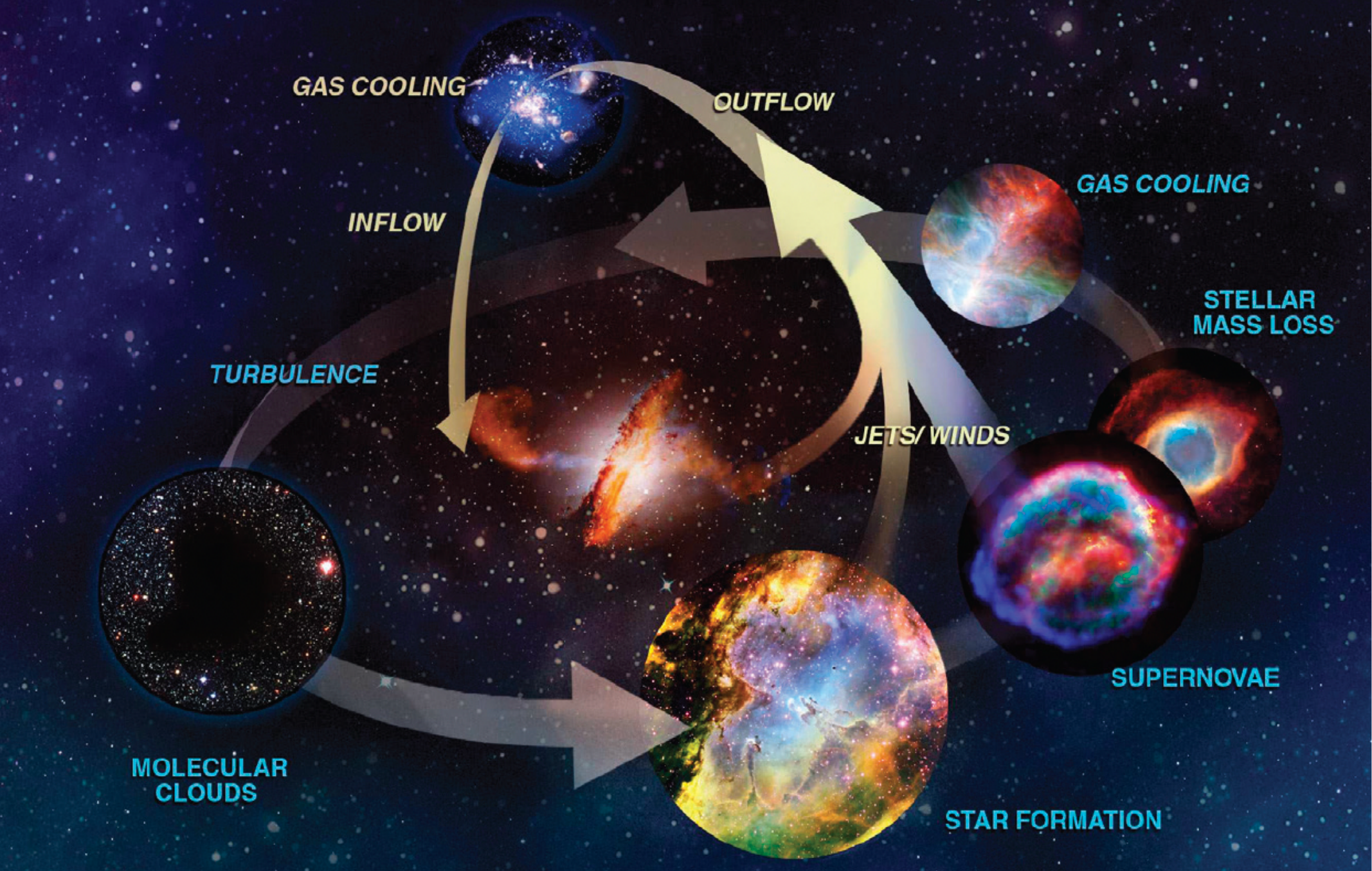}
\caption{\small Illustration of gas flows within galaxies, driven by both gravitational forces and feedback processes. AGN feedback, through jets and winds, plays a critical role in regulating these flows, impacting gas cooling, star formation, and the redistribution of material across the interstellar, circumgalactic, intracluster, and intergalactic media. This dynamic interaction is essential for understanding the lifecycle of galaxies and the broader cosmic ecosystem. Figure source: HABEX Report, The Habitable Exoplanet Observatory Study Team (see also {\bf Astro2020}).
\label{EvolutionCircle}}
\end{figure}

\begin{figure}[t]
\centering
\includegraphics[width=1\columnwidth]{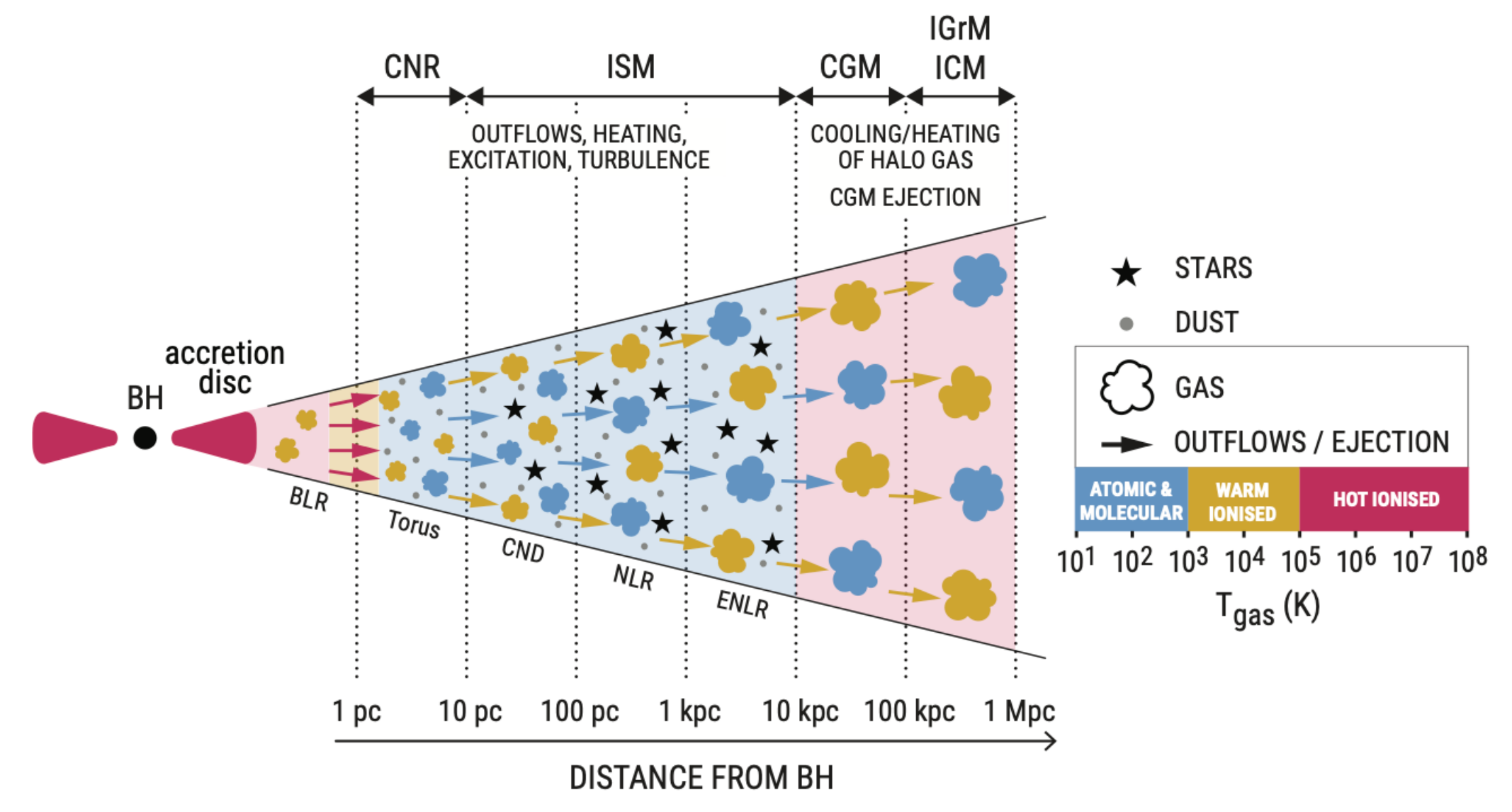}
\caption{\small A schematic diagram to illustrate the effects and spatial scales of AGN feedback and the role of outflows therein (adapted from \citealt{Harrison&RamosAlmeida2024} by the author's permission).
\label{MultiscaleOutflows}}
\end{figure}

As depicted in {\bf Astro2020} and illustrated in Figure~\ref{EvolutionCircle}, many aspects of galaxy formation and evolution can be attributed to results of a cosmic tug-of-war between feedback and gravitational collapse. Feedback here refers to how stars and black holes impact their surroundings through a broad set of energetic processes collectively. Specifically, the formation of galaxies is based on the cooling of the inflowing gas component, although as a manifestation of feedback, much of the gas is subsequently ejected back into the CGM by powerful galaxy-scale outflows. In theory, the flow of matter and energy within a galaxy determines the distribution of gas in the ISM and where and how stars and planetary systems form, leading to commonalities and differences between galaxies. However, understanding specific details in observation of the interplay between gravitational and feedback-driven processes such as outflows remains challenging, because it involves interconnected processes happening on a vast range of spatial and time scales as illustrated in Figures~\ref{EvolutionCircle} and \ref{MultiscaleOutflows}.

To solve this profoundly multi-scale problem affecting galaxy formation and evolution requires characterizing the feedback process from the {\it parsec-scale} impact radii of central accreting black holes to the {\it 10 kpc-scale} boundaries of galaxies and their {\it Mpc-scale} surroundings beyond. Specific observations in the spatially resolved way are pivotal to ascertain the role of feeding and feedback processes in shaping the cosmic ecosystem throughout such vast scales as highlighted by {\bf Astro2020}~D-Q2,3. Such observations include mapping the dynamics of gas flows in the ISM, CGM, IGM, and ICM, which are key to solving the mysteries of galaxy formation and evolution and are among the priority discovery areas of {\bf Astro2020} science panels

\section{Science Objectives}\label{sec2}

ISM gas is an essential component of galaxies responsible for both stellar growth (SF activity) and black hole growth (AGN activity). The understanding of how the feeding from and feedback upon a gaseous fuel supply achieve a balance between each other is crucial to deciphering the pathway of galaxy formation and evolution (see Figure~\ref{EvolutionCircle}), echoing the priority area of {\bf Astro2020} -- {\it Unveiling the Hidden Drivers of Galaxy Growth}. An important signature of the feedback in galaxy evolution is gas outflows of different phases over different physical scales, which are frequently observed in different types of galaxies throughout the nearby to distant universe. As illustrated in Figure~\ref{MultiscaleOutflows}, AGN feedback can affect different phases of the ISM gas with a wide range of temperatures. Moreover, corresponding feedback effects can occur over a gigantic range of physical scales from nuclear scales to galaxy cluster scales via gas outflows accompanying processes such as heating and turbulence.

\begin{figure}[!ht]
\centering
\includegraphics[width=0.8\columnwidth]{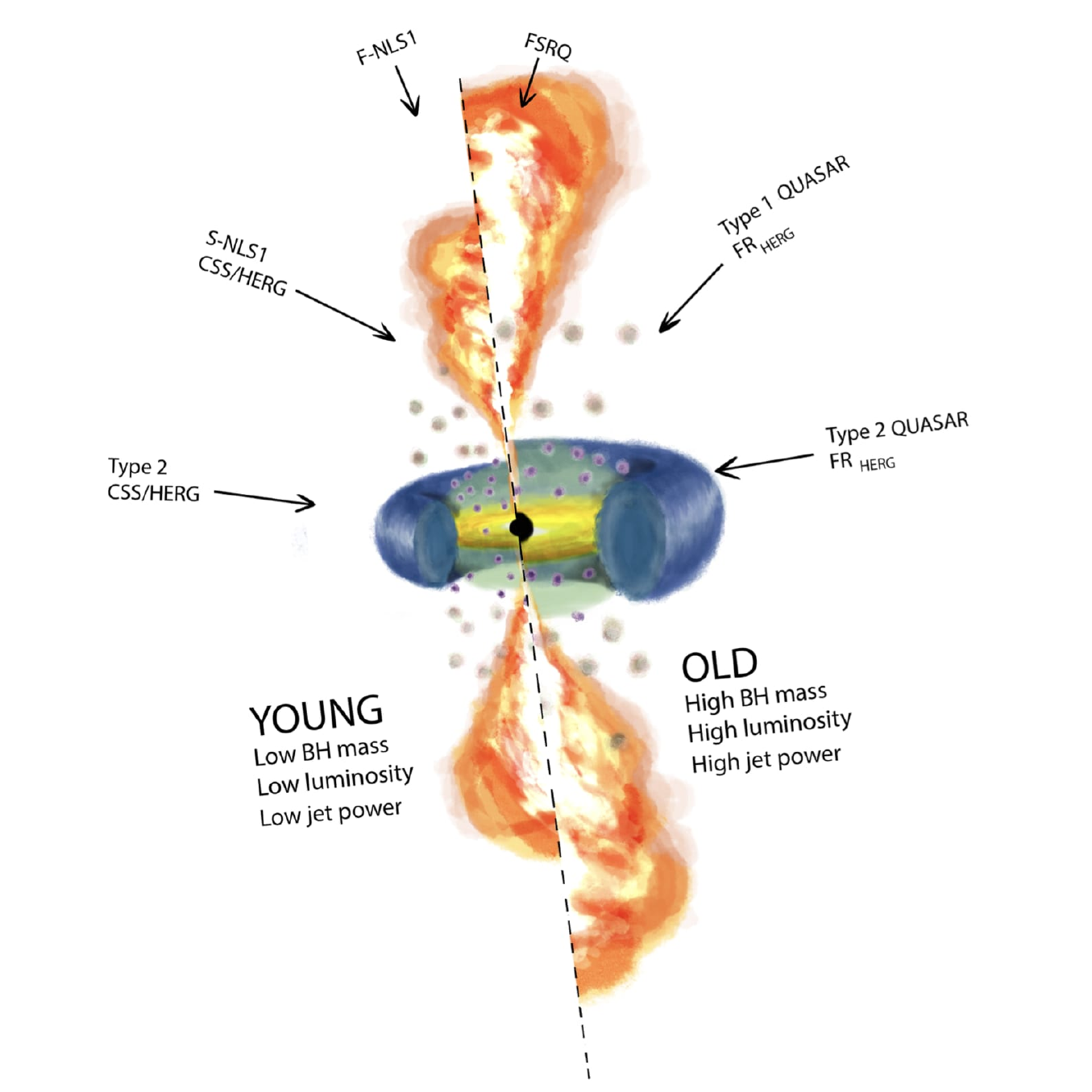}
\caption{\small A diagram highlighting the impact of AGN activity on the surrounding environment (adapted from \citealt{Berton.etal.2017} by the author's permission).
\label{Outflow}}
\end{figure}

AGN are widely regarded as the primary source of feedback affecting the evolution of massive galaxies and are regularly invoked in state-of-the-art cosmological simulations to inject energy into the ISM, regulate or quench star formation, and slow black hole growth. However, existing observations still dispute the effectiveness (i.e., positive or negative) and manner (kinetic or radiative mode) of AGN feedback regulating the evolutionary history of galaxies. Meanwhile, stellar feedback associated with star formation has been considered in low- and intermediate-mass galaxies, which drives ISM turbulence regulating the efficiency of star formation on scales from individual molecular clouds to entire galaxies. To better understand the role played by AGN feedback in galaxy evolution, it is essential to also consider the potential effects of stellar feedback. This includes examining how AGN activity is triggered and how it correlates with SF activity both before and after the onset of AGN activity.

Technologically, the above profoundly multi-scale and multi-phase problem drives investments in reaching high resolution and toward high sensitivity, as well as wide wavelength coverage as detailed in {\bf Astro2020}. Specifically, to determine the dominant feedback process (i.e., AGN feedback or stellar feedback) and their subsequent impact (as will be further discussed in Section~\ref{sec3}) at different physical scales (from $\sim$ {\it 1 pc} to {\it a Mpc}) in galaxy evolution, deciphering the launching and propagating of multi-phase gas outflows with spatially resolved observations by IFUs is indispensable. If the gas outflows in AGN host galaxies are primarily driven by their active black holes, we would expect the outflowing gas components to exhibit a trajectory directly linked to the central engine (i.e., the AGN accretion disk and the adjacent torus, on scales of $\sim$ {\bf 1--10 pc}; see Figures~\ref{MultiscaleOutflows} and \ref{Outflow}). Additionally, there should be a causal relationship between the strength of AGN activity (e.g., AGN luminosity from the accretion disk and the associated broad-line region) and the integrated strength of gas outflows (e.g., outflow mass rate and outflow energy; see review \citealt{McNamara&Nulsen2007} and \citealt{Fabian2012}).  The same applies to the stellar feedback but with a trajectory of outflowing gas components directly connected with star formation sites (i.e., molecular clouds on $\sim$ {\bf10--100 pc} scales), and causality between star formation properties (e.g., star formation rate; SFR) and outflow strength (see review \citealt{Veilleux.etal.2005}).

The IFU study of AGN components and outflows in the nearby universe with HWO would also help us better understand the role and impact of star formation and outflows in the high-redshift universe, whose observations are extremely challenging (e.g., the Little Red Dots; \citealt{Kocevski.etal.2023, Greene.etal.2024}). The dynamics, kinematics, and ionization structure of outflow, extending from the vicinity of the accretion disk to the outskirt of the host galaxy, are particularly poorly understood in high-redshift objects. Therefore, it is challenging to ascertain how their momentum and energy propagate into the galaxy and how they impact their environment. The HWO IFU study of the nearby universe allows us to investigate the relationship between AGN activity and recent star formation, as well as to examine how outflows influence and affect their host galaxies as a function of the fundamental properties of the AGN (e.g., black hole mass, accretion rate, SED). Such studies would provide a general framework for describing how outflows operate in AGN. The expansion of IFU outflow studies by the HWO in a larger population of AGN helps us better understand what physical factors govern the launch and duty cycle of AGN winds and how the outflow parameters scale with redshift.

Specific science objectives of this SCDD focusing on the spatially resolved study of the role played by feeding and feedback of growing black holes in galaxy evolution can be summarized as follows:
\begin{itemize}
\item Which is more dominant in triggering inflows and outflows through feedback: AGN activity or star formation, and how does it work? How is AGN activity triggered, and is it associated with circumnuclear star formation?
\item In AGN feedback, which mode is more influential: the momentum-driven (kinetic) mode or the energy-driven (radiative) mode?
\item What is the overall effect of AGN feedback on star formation -- does it predominantly inhibit or promote it (negative or positive feedback)?
\item Does AGN feedback operate similarly in the local universe and at high redshift, or does it vary across cosmic time?
\end{itemize}

\section{Physical Parameters}\label{sec3}

As specifically highlighted in {\bf Astro2020} (D-Q2,3): ``High-resolution, sensitive UV/optical/NIR spectral maps of galaxies, resolving H~{\small II} region scales out to z = 10 down to line sensitivities of $10^{-19}-10^{-20}\,\rm erg\,s^{-1}\,cm^{-2}$, coupled with spatially resolved sub-kpc X-ray, UV, and optical spectral maps of the CGM will establish gas kinematics and chemical imprints from the ISM to the CGM/ICM/IGM and connect small-scale feedback to gas properties on scales out to and beyond the virial radius'', ``sensitive rest-frame UV absorption spectroscopy with $\delta v < 10\,\rm km\,s^{-1}$ provides a powerful probe of the metal content and homogeneity in the warm diffuse CGM/ICM/IGM'', and ``to probe AGN winds across all the relevant ionization states and phases, high-throughput, high-resolution spectroscopy from the hard X-rays through the far-UV is needed''; ``deep, spatially resolved infrared and millimeter measurements are required to probe the molecular outflows and outflow dust content''. The science objectives of this SCDD are well aligned with the above statements, which require the spatially resolved (up to parsec scales to resolve the AGN torus) measurement of ionized and molecular gas content ($M$), as well as spatially resolved measurement of ionized and molecular gas kinematics ($v, \sigma$).

Specifically, to achieve the science objectives of this SCDD, we need to:
\begin{itemize}
\item Resolve and characterize the spatial distribution of ionized and cold/warm molecular gas, especially those in inflows/outflows.
\item Explore the spatial coupling and potential stratification of multi-phase inflows/outflows on different physical scales and their resolved and global correlations with AGN and/or SF activities.
\item Investigate whether corresponding outflows/jets induce shocks and/or fluctuations that trigger or suppress the formation of molecular clouds and, hence, new stars.
\end{itemize}

Accordingly, we need to measure the following physical parameters:
\begin{itemize}
\item Spatially resolved ionized  and molecular gas content ($M$, from the flux of gas emission)
\item Spatially resolved ionized  and molecular gas kinematics ($v, \sigma$ from emission line profile)
\item Spatially resolved ionized  and molecular gas outflow rate/energy distribution ($\dot{M}, \dot{E}$)
\item Observation and modeling of AGN strength and SF distribution ($L_{\rm AGN}$, $\lambda_{\rm AGN}$, SFR)
\end{itemize}

Now some following questions are:
\begin{itemize}
\item Which band or bands to observe to obtain the required physical parameters?
\item How to connect and interpret the above physical parameters in the framework of AGN feeding and feedback?
\item What is the adequate sample size for the analysis to draw convincing conclusions?
\end{itemize}

\begin{figure}[!ht]
\centering
\includegraphics[width=0.9\columnwidth]{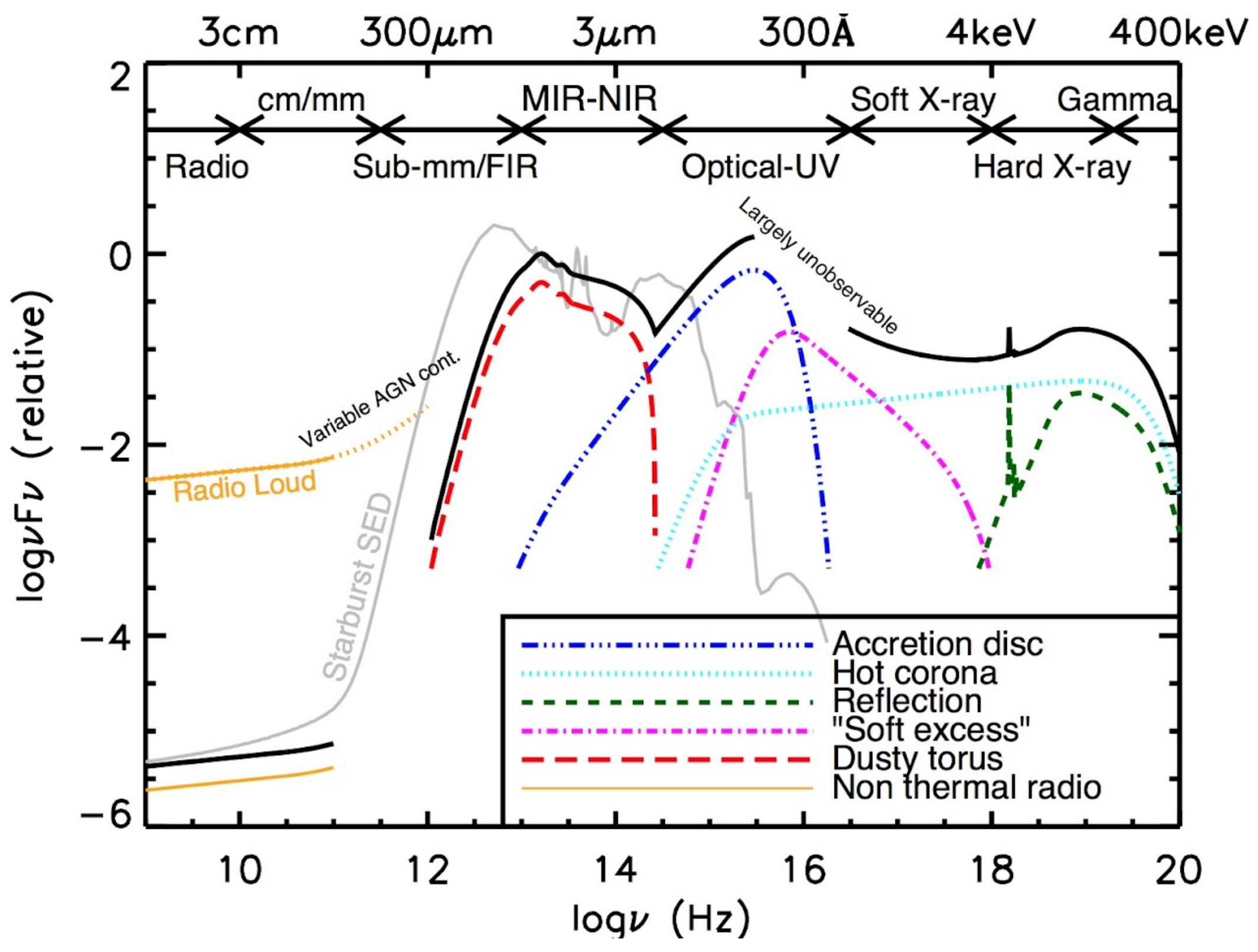}
\caption{\small A schematic representation of AGN spectral energy distribution (SED), loosely based on the observed SEDs of non-jetted quasars. The primary emission from the AGN accretion disk peaks in the UV region (adapted from \citealt{Harrison2014} by the author's permission).
\label{AGNSED}}
\end{figure}

\subsection{Which band or bands to observe to obtain the required physical parameters}\label{sec3.1}

\begin{figure*}[!ht]
\centering
\includegraphics[width=0.9\textwidth]{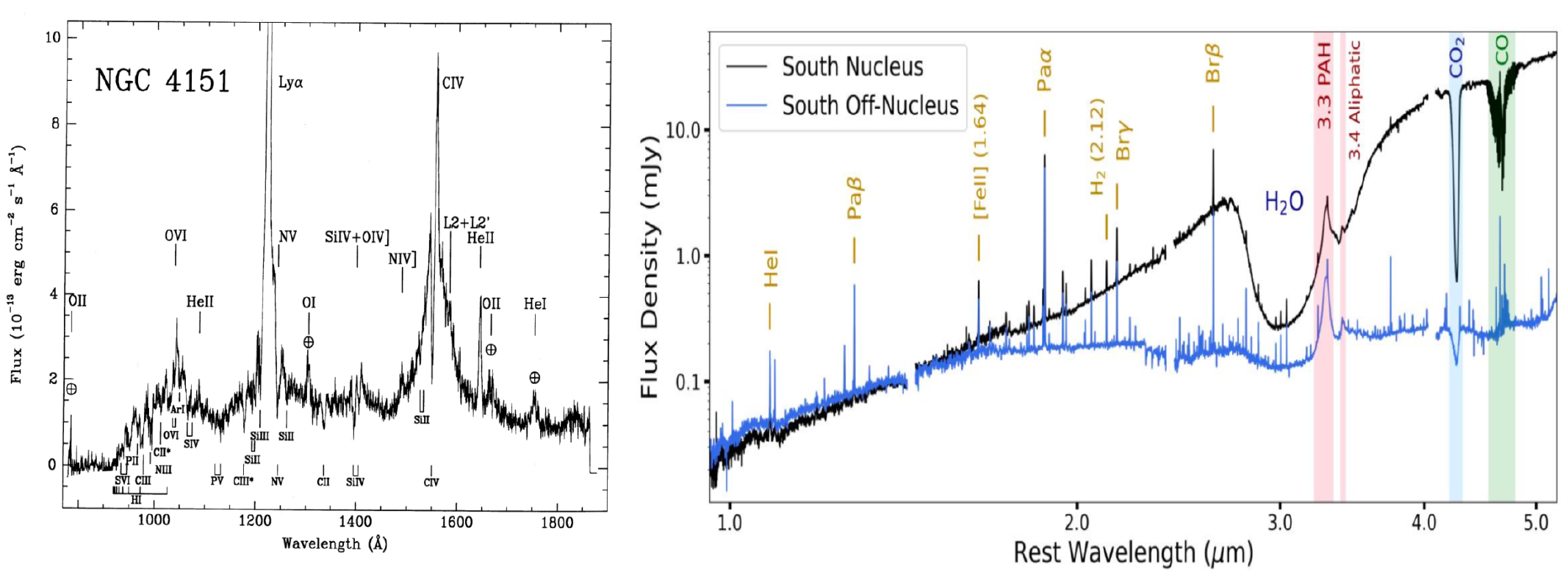}
\caption{\small Left: the UV spectrum of NGC 4151 (adapted from \citealt{Kriss.etal.1992} by the author's permission), where the identified emission and absorption features are marked above and below the data, respectively. The remaining airglow features in the spectrum are indicated with Earth symbols. Right: the NIR spectra of NGC 3256 with the marked emission and absorption features (adapted from \citealt{Bohn.etal.2024} by the author's permission).
\label{Spectra}}
\end{figure*}

To distinguish between different modes of AGN feedback and their effectiveness, the capability of UV band spectroscopy is indispensable. First, as illustrated in Figure~\ref{AGNSED},  UV emission is pivotal for constraining the nature and intensity of AGN activity and for breaking the degeneracy between AGN emission and emission associated with the host galaxy given the UV peaked emission of AGN accretion disk.  If there is recent star formation in the vicinity of the nucleus, the UV is also the best band to observe its signatures. Second, emission and absorption lines pertaining to UV spectra (see example in Figure~\ref{Spectra}) can provide important diagnostics of the physical properties of AGN-driven outflows and estimates of how much mass and energy they actually carry. Specifically, Lyman-alpha emission provides constraints on the mass and kinematics of ionized outflows (e.g., \citealt{Heckman.etal.2011, ArrigoniBattaia.etal.2016}).

\begin{figure*}[!ht]
\centering
\includegraphics[width=0.75\textwidth]{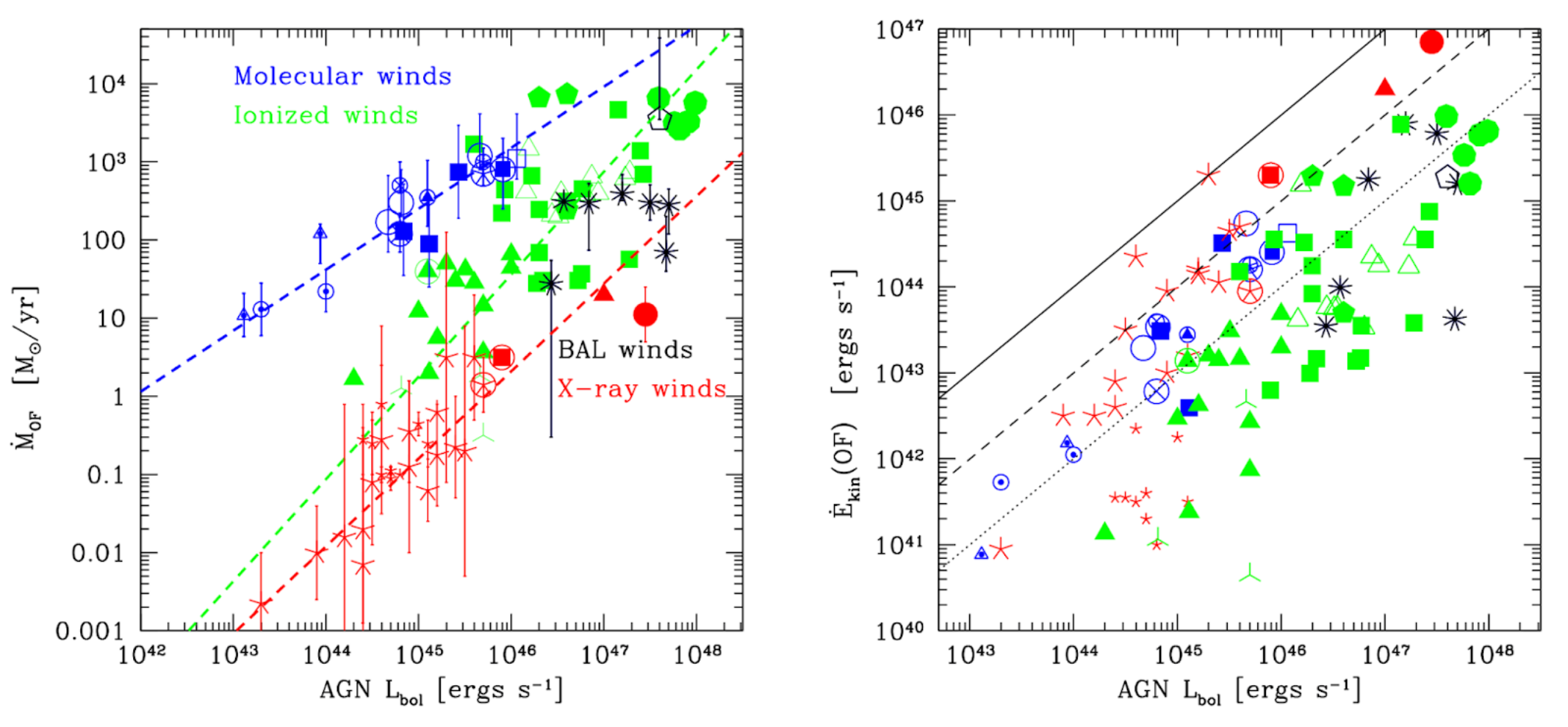}
\caption{\small Left: the mass outflow rate as a function of the AGN bolometric luminosity; Right: the outflow kinetic power as a function of the AGN bolometric luminosity (adapted from \citealt{Fiore.etal.2017} by the ESO's permission).
\label{CoupleOutflows}}
\end{figure*}

Meanwhile, the relatively dust-extinction-immune NIR spectroscopy is critical for robustly constraining models of the torus and then ascertaining the launching site of the AGN-driven gas outflows. The NIR spectrum also provides important information on stellar absorption features to trace stellar kinematics dominated by galaxy gravity, as well as complementary ionic and molecular absorption and emission lines to diagnose galaxy properties and estimate mass and energy carried by different phases of outflowing gas components, e.g., molecular/ionized hydrogen, CO, CO$_2$, H$_2$O, and polycyclic aromatic hydrocarbons (PAHs). Specifically, CO and H$_2$ features in the NIR spectrum provide constraints on the mass and kinematics of cold/warm molecular gas outflows (e.g., \citealt{Feruglio.etal.2010, Cicone.etal.2014, Cicone.etal.2017}), which are potentially linked to SF activities that can be traced by infrared PAH emission (e.g., \citealt{Kim.etal.2012, Shipley.etal.2016}).

\subsection{How to connect and interpret the above physical parameters in the framework of AGN feeding and feedback}\label{sec3.2}

Figure~\ref{CoupleOutflows} shows an example of existing quantitative studies on the role of AGN feedback in the co-evolution of black holes and galaxies, highlighting the global scaling relations between AGN properties, host galaxy characteristics, and outflows (\citealt{Fiore.etal.2017}). The correlations in Figure~\ref{CoupleOutflows} are consistent with the expectation that AGN feedback affects different phases of the ISM gas to different degrees, and the feedback occurs over a large range of spatial scales. Meanwhile, these correlations exhibit scatters, corresponding to different loading factors of the outflows, highlight the indispensability of the spatially resolved observations as will be detailed in Section~\ref{sec4}.

\subsection{What is the adequate sample size for the analysis to draw convincing conclusions}\label{sec3.3}

\begin{figure}[!ht]
\centering
\includegraphics[width=0.85\columnwidth]{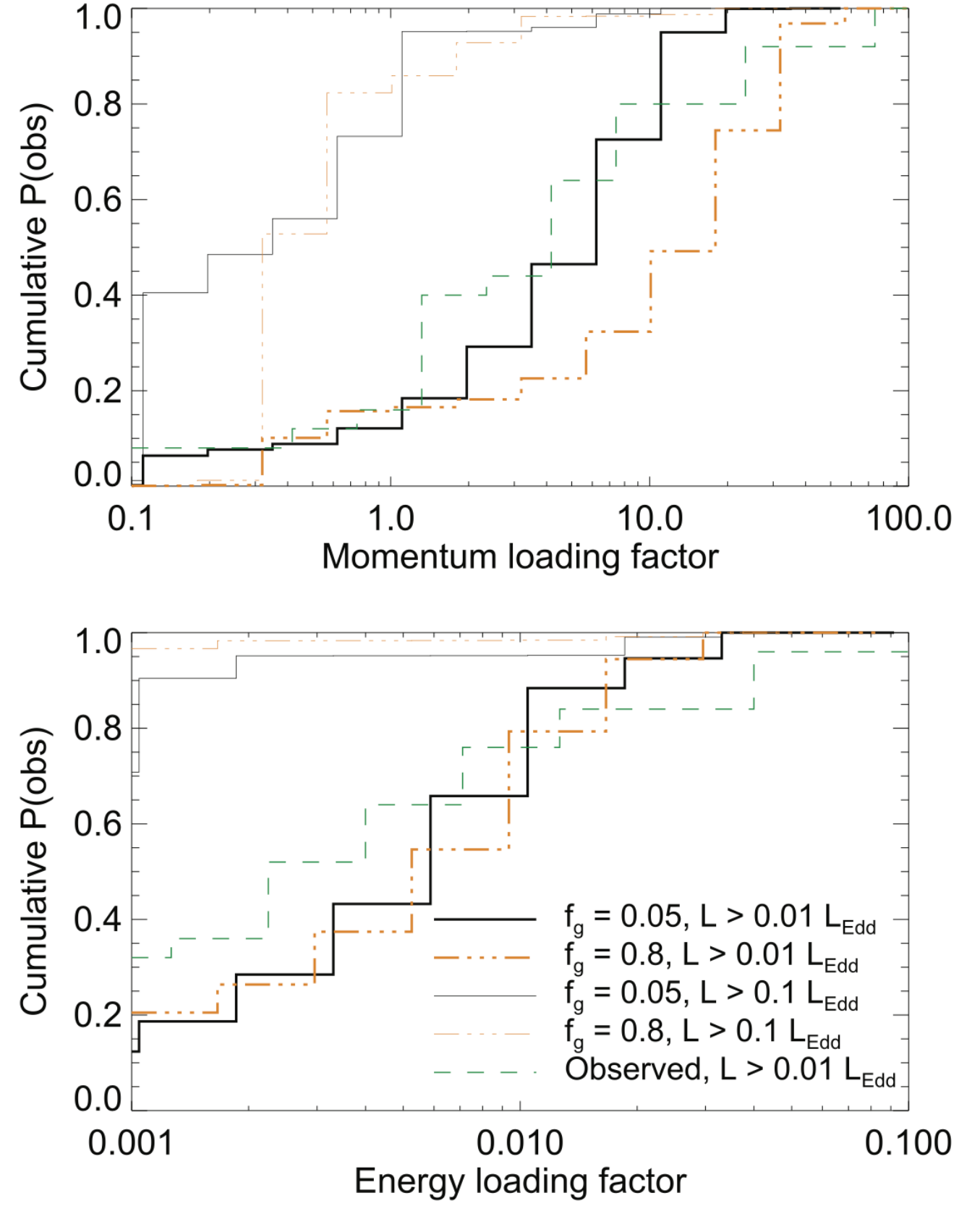}
\caption{\small Cumulative probability distribution of observable instantaneous momentum (top) and energy (bottom) loading factors in quasar models with different gas fractions ($f_{\rm g}$) and AGN luminosity. Only outflows with $R_{\rm out} < 10$ kpc are included. The green dashed line shows the loading factor distribution of observed molecular outflows in AGN with $L_{\rm AGN} > 0.01 L_{\rm Edd}$ (adapted from \citealt{Zubovas&Nardini2020} by the author's permission).
\label{Sampling}}
\end{figure}

As shown in Figure~\ref{Sampling} (the green dashed lines), both the observed momentum and energy loading factors of molecular outflows in AGN have a broad distribution, spanning over two to three orders of magnitude. Furthermore, AGN luminosity varies on a much shorter timescale ($\sim10^4-10^5$ yr; \citealt{Schawinski.etal.2015}) than the outflow lifecycle ($\sim 10^6$ yr; \citealt{Zubovas&Nardini2020}). These results mean that an outflow may persist for more than an order of magnitude longer than the AGN episode that drives it, leading to a substantial population of ‘fossil’ outflows (\citealt{King.etal.2011, Zubovas&Maskeliunas2023}). Namely, large-scale outflows away from the launch site are likely not physically related to the current AGN episode in observation, and hence, large sample statistics are required to ascertain the impact of AGN activity on galaxy evolution via outflows.

Therefore, to fully accomplish the science objectives of this SCDD as noted above, besides the spatially resolved observations and measurements of ionized and molecular gas outflow rate/energy ($\dot{M}, \dot{E}$), as well as AGN and SF strength ($L_{\rm AGN}$, $\lambda_{\rm AGN}$, SFR), we also need to carry out the spatially resolved analysis of the correlations between outflow parameters and AGN and/or SF parameters in an exhaustively statistical way. According to the distributions of the loading factor and duration of individual AGN episode and outflow, a factor of $\sim10-20$ for the sampling of different loading factors and a factor of $\sim50-100$ for the sampling of different evolution stages of outflows under different AGN episodes are desired for exhaustive statistics. Namely, it is ideal to have a sample of $\sim(10-20)\times(50-100)$, i.e., $\sim 500-2000$, galaxies for HWO IFU observations (see Table~\ref{tabobs}). High efficiency survey telescope UVEX (\citealt{Kulkarni.etal.2021}) is promising in the sample selection for further HWO IFU observations.

\begin{table*}[!ht]
\caption{Observation Parameters}
\smallskip
\begin{center}
{\small
\begin{tabular}{ccccc}  
\tableline
\tableline 
\noalign{\smallskip}
Observation Parameter & State of the Art & Incremental Progress & Substantial Progress& Major Progress\\
  &   & (Enhancing) & (Enabling) & (Breakthrough) \\
\noalign{\smallskip}
\tableline
\tableline 
\noalign{\smallskip}
Number of Galaxie & $\sim$ 100 galaxies & $\sim$ 500 with the & $\sim$ 1000 with the & $ \sim$ 2000 ($\sim$ 1\% of the \\
  & with $\sim 3\arcsec \times 3\arcsec$ NIR & coverage of $\sim 3\arcsec \times 3\arcsec$ & coverage of $\sim 6\arcsec \times 6\arcsec$ & galaxy number within \\
  & IFU coverage, 0 & each galaxy, 5 & each galaxy, 20 & 100\,Mpc from us) with \\
  & with complete IFU & with complete IFU & with complete IFU & the mosaic coverage of \\
  & coverage & coverage & coverage & $\sim 4 \times 10\arcsec \times 10\arcsec$ each, \\
  & & & & 50 with complete IFU \\
  & & & & coverage \\
\noalign{\smallskip}
\tableline 
\noalign{\smallskip}
Line flux sensitivity & $\sim10^{-17}\,\rm erg\,s^{-1}\,cm^{-2}$ & $\sim10^{-18}\,\rm erg\,s^{-1}\,cm^{-2}$ & $\sim10^{-19}\,\rm erg\,s^{-1}\,cm^{-2}$ & $\sim10^{-20}\,\rm erg\,s^{-1}\,cm^{-2}$ \\
\noalign{\smallskip}
\tableline 
\noalign{\smallskip}
Velocity dispersion & $\sim 50\,\rm km\,s^{-1}$ & $\sim 30\,\rm km\,s^{-1}$ & $\sim 20\,\rm km\,s^{-1}$ & $\sim 10\,\rm km\,s^{-1}$ \\
\noalign{\smallskip}
\tableline 
\noalign{\smallskip}
$L_{\rm AGN}$ & $\sim 10^{43}-10^{48}\,\rm erg\,s^{-1}$ & $\sim 10^{42}-10^{48}\,\rm erg\,s^{-1}$ & $\sim 10^{41}-10^{48}\,\rm erg\,s^{-1}$ & $\sim 10^{40}-10^{48}\,\rm erg\,s^{-1}$ \\
\noalign{\smallskip}
\tableline 
\tableline 
\end{tabular}
}\label{tabobs}
\end{center}
\end{table*}

\section{Description of Observations: Why HWO}\label{sec4}
\begin{itemize}
\item HWO's capabilities will enable the spatially resolved analysis of multi-phase outflows and their connection to AGN activity across diverse galaxy types.
\item Existing facilities lack the required combination of high-throughput UV and NIR IFU capabilities with sufficient spatial resolution and sensitivity.
\end{itemize}

First of all, to accomplish the science objectives of this SCDD, we need to have the spatially resolved maps (with a spatial resolution of up to {\bf parsec-scale} torus sizes) of ionized and molecular gas content, as well as their kinematic information. The spatially resolved IFU (not multi-object spectrograph as mentioned later) observations will enable us to distinguish the gas content in inflows/outflows (according to their kinematics with the velocity dispersion down to $10\,\rm km\,s^{-1}$, e.g., \citealt{Wisnioski.etal.2015}) and the launching sites of outflows (AGN or SF regions) to trace the evolutionary pathway of gas inflows/outflows.

Specifically, we need to perform such analyses for different phases of gas outflows, i.e., ionized gas outflows (requires UV/NIR spectra; see Figure~\ref{Spectra}) and cold/warm molecular gas outflows (requires NIR spectra for CO$2$, CO, and H$2$ features; right panel in Figure~\ref{Spectra}). We also need to include the analysis of the spatial coupling and potential stratification of different phases of gas content to estimate the overall mass and energy loading carried by gas inflows/outflows. In addition, metallicity provides a quantitative measure of the enrichment history of different gas reservoirs. Metallicity measurements rely on observations of weak, narrow diagnostic lines that appear primarily in the rest-frame UV and optical. Particularly, sensitive rest-frame UV absorption spectroscopy with $\delta v < 10\,\rm km\,s^{-1}$ provides a powerful probe of the metal content and homogeneity in the warm diffuse CGM/ICM/IGM ({\bf Astro2020} D-Q2b).

The spatially resolved analysis is also required to diagnose the potential outflow/jet-induced localized shocks and/or fluctuations that trigger or suppress the formation of molecular clouds and, hence, new stars. Jets (especially on compact sources) are known to ionize the gas inside the jet lobe and in the surrounding medium via shocks. These shocks could also heat up and increase the turbulence in the SF-surrounding gas. Also, fast shocks (and high radiation levels from the AGN) could destroy PAHs, whose emission can be used to quantify SFR in powerful AGN. Some radio sources, however, show UV emission aligned with the radio jet, which could suggest jet-induced SF in old elliptical galaxies.

\begin{table*}[!ht]
\caption{Instrument Parameters}
\smallskip
\begin{center}
{\small
\begin{tabular}{ccccc}  
\tableline
\tableline 
\noalign{\smallskip}
Instrument Parameter & State of the Art & Incremental Progress & Substantial Progress& Major Progress\\
  & (IFU observation)  & (Enhancing) & (Enabling) & (Breakthrough) \\
\noalign{\smallskip}
\tableline
\tableline 
\noalign{\smallskip}
Type (imaging, & NIR IFU with the &   & NIR \& UV IFUs with & NIR \& UV IFUs with \\
spectroscopy, etc.) & FOV of $\sim 3\arcsec \times 3\arcsec$ &   & the FOV of $\sim 6\arcsec \times 6\arcsec$ & the FOV of $\sim 10\arcsec \times 10\arcsec$ \\
\noalign{\smallskip}
\tableline 
\noalign{\smallskip}
 Wavelength Range & NIR $\sim 1-5\,\rm\mu m$ &  & UV  $\sim$ 100 -- 400 nm; & UV  $\sim$ 100 -- 900 nm; \\
  &  &  & NIR $\sim 1-3\,\rm\mu m$ & NIR $\sim 1-5\,\rm\mu m$ \\
\noalign{\smallskip}
\tableline 
\noalign{\smallskip}
 Spatial resolution & NIR $\sim 0\farcs02 - 0\farcs15$ &  & NIR $\sim 0\farcs04$ at 1~$\mu m$ & NIR $\sim 0\farcs03$ at 1~$\mu m$ \\
 (Diffraction limited &  &  & \& UV $\sim 0\farcs004$ at & \& UV $\sim 0\farcs003$ at \\
 FWHM of PSF) &  &  & 100 nm for 6 m HWO & 100 nm for 8 m HWO \\
\noalign{\smallskip}
\tableline 
\noalign{\smallskip}
 Spectral resolution & R $\approx$1500 -- 4000 &  & R$>$3000 for NIR & R$>$5000 for NIR \\
 (R = $\lambda/\delta\lambda$) &   & & R$>$30000 for UV  & R$>$50000 for UV  \\
\noalign{\smallskip}
\tableline 
Limiting sensitivity & $\sim10^{-17}\,\rm erg\,s^{-1}\,cm^{-2}$ &  & $\sim10^{-19}\,\rm erg\,s^{-1}\,cm^{-2}$ & $\sim10^{-20}\,\rm erg\,s^{-1}\,cm^{-2}$ \\
\noalign{\smallskip}
\tableline 
\tableline 
\end{tabular}
}\label{tabinst}
\end{center}
\end{table*}

Besides the outflow properties, UV spectra of galaxies also provide a great constraint on AGN intrinsic SED and reliable ISM diagnostics, as well as on signatures of the presence of young stars. The available UV spectra taken by HST slits have the limited sensitivity of $\sim10^{-17}\,\rm erg\,s^{-1}\,cm^{-2}$ , while there is currently no UV IFU available for the spatially-resolved spectroscopy survey as also requested in {\bf Astro2020} Science White Paper (\citealt{James.etal.2019}). Like the NIRSpec Multi-Object Spectrograph (MOS; with the FOV of $3\farcm6\times3\farcm4$) on the JWST, the LUVOIR proposed to have an UV MOS, i.e., LUMOS, with the FOV of $2\arcmin\times2\arcmin$ (R $\approx$ 500--50000). However, unlike the JWST with an additional NIRSpec IFU,  there will be still no UV IFU on the LUMOS. An UV IFU on the HWO is indispensable for our science case as MOS is best for observing multiple targets close to each other (e.g., a cluster of galaxies), while IFU is best for studying a single target with extended flux distribution. Specifically, MOS provides spectra for individual points while losing most spatial information given its essence of discrete shutters, whereas IFU gives a detailed spatial map of the emission across the target area.

Meanwhile, although with competitive spatial resolutions (PSF FWHM: $\sim 0\farcs02 - 0\farcs15$ for JWST/NIRSpec and $\sim 0\farcs1 - 0\farcs25$ for ground-based instruments with AO) and spectral resolutions (R $\approx$ 1500--4000), the existing NIR IFUs (e.g., JWST/NIRSpec IFU; Keck/OSIRIS; VLT/KMOS; Gemini/NIFS) overall have limited FOVs ($\sim 3\arcsec \times 3\arcsec$) and hence are not effective for mapping nearby galaxies, which are ideal targets for the spatially resolved analysis. Moreover, while free of atmospheric absorption, UV observations from space can still be affected by interstellar dust extinction, especially for observations targeting galactic centers. Therefore, simultaneous NIR observations are critical to compensate for the influence of dust extinction on UV observations while also providing extra constraints in the SED modeling of AGN properties. 

Considering the state-of-the-art capability of currently existing NIR IFUs, as well as to more efficiently map and resolve a galaxy and accurately measure the kinematics of UV/NIR emission and absorption features, UV and NIR IFU spectroscopies with $\sim 6\arcsec \times 6\arcsec$ or better $\sim 10\arcsec \times 10\arcsec$ FOV, $\sim 0\farcs01$ or better PSF size (with the resolved physical scale $<$ 10 pc or better out to z = 0.1, i.e., 400 Mpc, and $<$ 1 kpc at all redshifts), R $\approx$ 3000 -- 30000 spectral resolution ($\delta v \approx 10-100\,\rm km\,s^{-1}$), and $\sim10^{-19}\,\rm erg\,s^{-1}\,cm^{-2}$ or higher sensitivity are required (see Table~\ref{tabinst}). Furthermore, a diffraction-limited and stable PSF of the IFU spectroscopy is very important for (1) achieving the best spatial resolution and (2) removing the even minor but potential smearing effect that will affect the accuracy of emission line width measurement and hence the outflow properties.

{\bf Acknowledgements.} L.Z. acknowledges grant support from the Space Telescope Science Institute (ID: JWST-GO-01670; JWST-GO-03535; JWST-GO-04225; JWST-GO-04972). T.G. acknowledges support from ARC Discovery Project DP210101945. I.M. acknowledges financial support from the State Agency for Research of the Spanish MCIU, through the ``Center of Excellence Severo Ochoa'' award to the Instituto de Astrof\'{i}sica de Andaluc\'{i}a (CEX2021-001131-S), and through PID2022-140871NB-C21.

\bibliography{zhang.bib}

\end{document}